# Phosphorylation potential and chemical fluxes govern the biological performance of multiple PdP cycles

Teng Wang[1,3], Chenzi Jin[2,3], Fangting Li[2,3*]

1 School of Life Sciences, Peking University, Beijing 100871, China.
2 School of Physics, Peking University, Beijing 100871, China.
3 Center for Quantitative Biology, Peking University, Beijing 100871, China.
* Correspondence: lft@pku.edu.cn

## Abstract

Fission yeast G2/M transition is regulated by a biochemical reaction networks which contains four components: Cdc13, Cdc2, Wee1, and Cdc25. This circuit is characterized by the ultrasensitive responses of Wee1 or Cdc25 to Cdc13/Cdc2 activity, and the bistability of Cdc2 activation. Previous work has shown that this bistability is governed by phosphorylation energy. In this article, we developed the kinetic model of this circuit and conducted further thermodynamic analysis on the role of phosphorylation energy ($\Delta G$). We showed that $\Delta G$ level shapes the response curves of Wee1 or Cdc25 to Cdc2 and governs the intrinsic noise level of Cdc2 activity. More importantly, the mutually antagonistic chemical fluxes around the PdP cycles in G2/M circuit were shown to act as a stabilizer of Cdc2 activity against $\Delta G$ fluctuations. These results suggest the fundamental role of free energy and chemical fluxes on the sensitivity, bistability and robustness of G2/M transition.

## Introduction

The onset of mitosis is one of the major events during eukaryotic cell cycle (1). The irreversibility of this transition from G2 phase to M phase, which means that the cell persists in the M state after the driving factors is withdrawn, is crucial for cells because it ensures the unidirectional of cell cycle and provides the robustness against cellular noises (2, 3). Eukaryotic G2/M transition is subject to a universal control mechanism (1). For fission yeast, in the center of this mechanism is Cdc13/Cdc2 kinase complex, the activation of which brings about the entry into M phase (1, 4). One inhibitory kinase, Wee1, suppressed the activity of Cdc2 by phosphorylation on its Tyr15 site (5), whereas Cdc25, a phosphatase, removes this inhibitory phosphate and thus activates Cdc13/Cdc2 (6). The activities of both Wee1 and Cdc25 are regulated by multisite phosphorylation catalyzed by Cdc2. Cdc25 is activated by Cdc2 (7) and Wee1 is inhibited (8). Therefore, Cdc2 and Cdc25 comprise a positive feedback while Cdc2 and Wee1 comprise a double-negative feedback.

Experiments using *Xenopus laevis* egg extracts have detected the bistability of this controlling mechanism and conformed that cyclin B-driving Cdc2 activation is switch-like and exhibits hysteresis (9, 10). In customary view, such bistability originates from two sources. First, the two feedbacks in G2/M circuit constitute a bistable trigger (2, 9-12). When the feedback loops are compromised, the activation of Cdc2 becomes gradual and oscillations in Cdc2 activity is damped (13). Second, the responses of both Wee1 and Cdc25 to Cdk1 are ultrasensitive (14, 15). One of the contributors of such sensitivity is the multisite phosphorylation of Wee1 or Cdc25 (14, 15). At least five sites of Wee1 or Cdc25 are phosphorylated when cell enters M phase (16, 17), although the details about phosphorylation network among these sites are still unclear. Feedbacks and ultrasensitivity, combined together, create a toggle switch with two discrete state: G2 state (inactive Cdk1, Cdc25 and active Wee1) and M state (active Cdk1, Cdc25 and inactive Wee1) (9, 18).

However, the role of ATP consumption in G2/M is still to be understood. Accompanying the phosphorylation and dephosphorylation (PdP) cycles of these three components, ATP hydrolysis provides not only phosphate as modifications, but also free energy. The chemical basis for such free energy release is that living cell is sustained away from equilibrium with high concentration of ATP and relatively low concentrations of ADP and Pi. In normal cells, the phosphorylation potential ($\Delta G$), which is defined as the amount of free energy released from one mole of ATP hydrolyzed to ADP and Pi, is within the range of 50~60 kJ/mol (19). Such non-zero phosphorylation potential drives the chemical fluxes around PdP cycles, and free energy is consumed in the form of heat dissipation (20). In recent years, a NESS (nonequilibrium steady state) theory has been well established and throws light on the relation between the nonlinear behaviors and the nonequilibrium nature of biochemical reactions (20). In their theoretical works focusing on some examples of biochemical reaction systems, free energy input governs the biological functions such as zero-order ultrasensitivity (21), ligand-receptor specificity (22), and oscillations (23). Following these findings, it is natural to ask how phosphorylation potential level influences the behaviors of G2/M system.

Recently, we utilized fission yeast nuclear extracts and systematically explored the role of $\Delta G$ during Cdk1 activation (24). It is observed that increasing $\Delta G$ drives the activation of Cdc2, and as with Cdc13, this $\Delta G$-driving Cdc2 activation is accompanied by hysteresis. Besides, Cdc13-driving Cdc2 activation is only possible when $\Delta G$ is large enough. When the extract is incubated at 30°C for 4h to let $\Delta G$ fall to lower than 46 kJ/mol, adding Cdc13 to the extract fails to activate Cdk1. These results emphasize the fundamental importance of free energy input to M entrance and renew the traditional picture of G2/M transition: bistability or irreversibility of this system strongly depends on the level of phosphorylation potential.

However, several questions regarding the role of free energy input of G2/M system remain to be answered. First, how does $\Delta G$ level influence the responses of Wee1 and Cdc25 to Cdc2? For PdP cycles operating at zero-order zone, it has been shown that $\Delta G$ determines the sensitivity of the response and raising $\Delta G$ increases the steepness of the response curves (21). However, zero-order ultrasensitivity contributes little to the ultrasensitivity of Wee1 or Cdc25 because Cdc2 is operating way from saturation (14, 15). The role of $\Delta G$ in ultrasensitivity generated from multisite phosphorylation is little known yet. Second, how should the directivity of chemical fluxes on the PdP cycles of G2/M system been understood? How would the dynamics of this system be changed if one of these fluxes was reversed? To investigate these questions, we build a simple model describing G2/M circuit as a set of PdP cycles. By admitting the reversibility of the enzymatic reactions (25), this model allows both kinetic and thermodynamic analysis of the Wee1/Cdc2/Cdc25 circuit.

## Model and results

**ΔG shapes the response curve of the multiple PdP cycles.** Before turning to the whole model of G2/M regulatory network which includes two feedbacks, we first investigate the steady-state responses of Wee1 or Cdc25 activity to the active concentration of Cdc13/Cdc2. The multi-site phosphorylation process of Wee1 or Cdc25 catalyzed by active Cdc13/Cdc2 is described as a general scheme which is composed of a sequence of PdP cycles, as is shown in Fig. 1A. Here, we have assumed that the kinase and phosphatase act according to a distributive mechanism, which means the enzymes release the substrate after each step of modification (26). Every individual PdP cycle is coupled with the hydrolysis of one ATP molecule and composed of two irreversible reactions

$$S_n + K + ATP \underset{b_{n+1}}{\overset{a_n}{\rightleftarrows}} S_{n+1} + K + ADP, \quad [1]$$

$$S_{n+1} + P \underset{d_n}{\overset{c_{n+1}}{\rightleftarrows}} S_n + P + Pi, \quad [2]$$

where $n=0, 1......, N-1$ and $a_n, b_{n+1}, c_{n+1}$, and $d_n$ are the rate constants for the $n+1$ th PdP cycle. Eq. **[1]** describes the phosphorylation process of substrate ($S$) catalyzed by kinase ($K$), whereas eq. **[2]** describes the dephosphorylation catalyzed by phosphatase ($P$). The subscript $n$ of $S$ represents that $n$ sites on substrates have been phosphorylated. According to the NESS theory, the intracellular phosphorylation potential $\Delta G$ is directly related to the quotient $\gamma = \dfrac{a_n c_{n+1}[ATP]}{b_{n+1} d_n [ADP][Pi]}$ with

$\Delta G = RT \ln \gamma$ where $R$ is the ideal gas constant and $T$ is the Kelvin temperature (20-22). $RT$ equals about 2.48 kJ/mol at room temperature.

For the sequential phosphorylation mechanism, under which phosphorylation sites are modified following a strict order, we assume that the rate constants are independent of the phosphorylation states of the substrates ($a_n=a, b_n=b, c_n=c, d_n=d$) (26). According to Eq. **[1]** and **[2]**, at steady state, the ratio between $[S_{n+1}]$ and $[S_n]$, which is written as $\sigma_n$, becomes

$$\sigma = \frac{(\gamma \tilde{b}\tilde{d}/c)\, x + \tilde{d}}{\tilde{b} x + c}, \quad [3]$$

where $\tilde{b} = b[ADP]$, $\tilde{d} = d[Pi]$. Here $x$ denotes the ratio $[K]/[P]$ and is regarded as the stimuli of the PdP cycles. It is also assumed that the activity of substrate (Wee1 or Cdc25) only turns over at the last step of phosphorylation. In this case, the steady-state fraction of totally phosphorylated substrate $y_N$ is defined as the output. It can be easily obtained that $y_N$ is related to $\sigma$ with $y_N = \sigma^N(\sigma-1)/(\sigma^{N+1}-1)$.

This distributive and sequential mechanism produces sigmoid dose-response curves (Figs. 1B-D). When $x$ increases continuously, $y_N$ approaches one limit $y_N^{\max}$, the value of which is determined by $\lim_{x \to +\infty} \sigma = \gamma \tilde{d}/c$. To give rise to adequate downstream changes, a proper $y_N^{\max}$ in living cell is required to be nearly 1. Fig. 1E shows how $N$ and phosphorylation potential determines the value of $y_N^{\max}$. The sufficient inactivation of Wee1 or activation of Cdc24 is only possible when $\Delta G$ is higher than a threshold (~31 kJ/mol). Beyond this threshold, the number of phosphorylation sites has little influence on $y_N^{\max}$. There exist a 'dead district' (red dashed box) where the PdP cycles give hardly no responses, no matter how large $x$ becomes. For cellular molecular modules that transfer signals through phosphorylation modification, the information flux might be completely blocked if $N$ and $\Delta G$ falls in this district. The second index, $x_{half}$, under which $y_N$ equals to $0.5 y_N^{\max}$, characterizes the strength of stimuli to produce a half-maximum response. A over-high $x_{half}$ makes this module useless for cell due to the incapability to synthesize infinite kinase molecules. Raising $\Delta G$ to a range of 53.2~64.5 kJ/mol reduces $x_{half}$ to a proper order of magnitude of 1~10 (Fig. 1F). However, when $\Delta G$ increases further from 64.5 kJ/mol, the resulting over-low $x_{half}$ might lead to the intolerance to intracellular noises.

Next, we define a coefficient, $R_{0.5}$, as the ratio of $x$ when $y_N = 0.1 y_N^{\max}$ to the $x$ when $y_N = 0.5 y_N^{\max}$. $R_{0.5}$ characterizes the sharpness of the response curve (Fig. 1D), and with higher $R_{0.5}$, the output exhibits higher sensitivity to stimuli and better signal-to-noise ratio. Fig. 1G shows how $R_{0.5}$ depends on both $N$ and phosphorylation potential. Generally, increasing the number of phosphorylation sites or $\Delta G$ makes the response more sensitive. However, within the range $N$=1~6 or $\Delta G$=25~30 kJ/mol, $R_{0.5}$ raises dramatically. There exist a plateau area at the right-top corner where $R_{0.5}$ is close to its maximal (~0.546) and quite robust to $N$ or $\Delta G$ fluctuations. Interestingly, the physiological levels of $\Delta G$ is shown to be in this range. The steepness of the curves in Fig. 1D can also be characterized by the slope at $x/x_{half}$=0.5, which is refered as $n_v$. As is shown in Fig. 1H, $n_v$ exhibits the same tendency as $R_{0.5}$ when $N$ or $\Delta G$ changes. These four indexes combined together, we predict that the suitable range of phosphorylation potential is 53.2~64.5 kJ/mol where the normal function of PdP module is ensured.

If the sites are modified in a completely random manner, the rate constants will depend on the number of the available sites ($a_n=(N-n)a$, $b_n=nb$, $c_n=nc$, $d_n=(N-n)d$). In this case, $\sigma_n$ becomes $\sigma(N-n)/(n+1)$ where $\sigma$ is defined in Eq. [3] and $y_N$ becomes $\sigma^N/(1+\sigma)^N$ (26). Compared to sequential mechanism, under random mechanism, $y_N^{max}$ and $x_{half}$ show the similar dependence on $\Delta G$ (Figs. S2A, B). However, the sequential mechanism generates steeper response curves and a stronger dependence of $R_{0.5}$ or $n_v$ on $\Delta G$ (Figs. S2C-E). If the substrate is modified randomly, the number of phosphorylation sites seems to be the main factor determining the sensitivity of the curves.

In our model, the enzymatic addition or removal of phosphate on substrate is treated as elementary reactions. Another mathematical description of the multi-site phosphorylation process based on elementary steps of enzyme-substrate association and catalysis is also considered. As is shown in *supporting information*, with an assumption of rapid-equilibrium of enzyme-substrate binding, these two models generate the similar results. However, the basic models we have discussed here have fewer parameters and states, which provide convenience for a kinetic analysis.

**Phosphorylation energy determines the bistability of G2/M transition**. Once completely phosphorylated, Wee1 loses the activity of inhibiting Cdc2, while totally phosphorylated Cdc25 removes the inhibitory phosphate on Cdc2. Such reaction network is shown in Fig. 2A. The central kinase, Cdc13/Cdc2, undergoes a PdP cycle as follows:

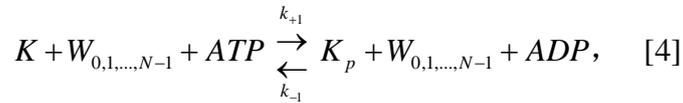

$$K + W_{0,1,...,N-1} + ATP \underset{k_{-1}}{\overset{k_{+1}}{\rightleftarrows}} K_p + W_{0,1,...,N-1} + ADP, \quad [4]$$

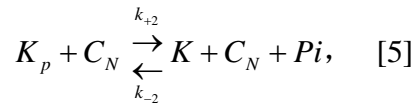

$$K_p + C_N \underset{k_{-2}}{\overset{k_{+2}}{\rightleftarrows}} K + C_N + Pi, \quad [5]$$

where $k_{+1}$, $k_{-1}$, $k_{+2}$, $k_{-2}$ are the rate constants. $K_p$ is the phosphorylated form of central kinase. $W_{0,1,...,N-1}$ represents the Wee1 molecules without phosphate modifications or partially phosphorylated, while $C_N$ represents the active form of Cdc25, which has been fully phosphorylated. Around this PdP cycle, the potential $\Delta G$ is similarly related to the rate constants of Eq. [4] and [5] with $\gamma = \dfrac{k_{+1}k_{+2}[ATP]}{k_{-1}k_{-2}[ADP][Pi]}$ and

$\Delta G = RT \ln \gamma$. Then the ODEs describing the dynamics of this system could be easily obtained (*supporting information*). The central ODE regarding the activation and deactivation of Cdc13/Cdc2 is as follows:

$$\frac{dz}{d\tau} = -\frac{\gamma \tilde{k}_{-1} \tilde{k}_{-2}}{k_{+2}} z f_w + \tilde{k}_{-1}(1-z)f_w + \theta k_{+2}(1-z)f_C - \theta \tilde{k}_{-2} z f_C, \quad [6]$$

where $\tilde{k}_{-1} = k_{-1}[ADP]$, $\tilde{k}_{-2} = k_{-2}[Pi]$, $\theta = [Cdc25]^{total}/[Wee1]^{total}$ and $\tau = t[Wee1]^{total}$. $z, f_w, f_c$ represents the active fractions of Cdc13/Cdc2, Wee1 and Cdc25, respectively.

When the system reaches steady state, $f_w = 1 - y_N^{wee1}$ and $f_c = y_N^{Cdc25}$ where the stimulus depend on the product of $z$ and $T$ ($T = [Cdc13/Cdc2]^{total}/[Wee1]^{total}$, see *supporting information*). Then we can calculate the fixed points (the steady-state $x$) of the system and analyze their stability. Increasing $T$ or $\Delta G$ induce the saddle-node bifurcation of the system, as is shown in Figs. 2B and C, suggesting that both $T$ and $\Delta G$ function as the driving factor of the bistable switch behavior during G2/M transition. It is noteworthy that, to make the threshold of $T$ reasonable for living cells, it's necessary to assign a basal activity to Cdc25, which is small but trigger the activation of Cdc2 (*supporting information*). Without this trigger, the threshold of T for entering M phase would become remarkable high. In our model, we assume a small amount (10% of $T$) of a certain kinase which together with Cdc13/Cdc2 activate Cdc25. This Cdc25-specific kinase can be viewed as a new species of proteins in our network or the 'leakage' of inactive Cdc13/Cdc2. Fig. 2D shows how these two factors together determined the steady states of the system and the bistability of G2/M trajectory. Increasing $\Delta G$ deduce the $T$ threshold, while increasing $T$ brings down the free energy requirement for M entrance.

**Phosphorylation energy induces critical slowing down.** Theoretical studies have suggested that a system tend to exhibit 'critical slowing down' when approaching bifurcation point (27, 28). The symptoms of 'critical slowing down', including slow recovery rate and increase in autocorrelation and variance, are often used as the early warning-signals for critical transition (28).

To illustrate this phenomenon in $\Delta G$-driving G2/M transition, we expose our theoretical model to stochastic perturbations of $z$ under different levels of $\Delta G$. When $T=1.4$, the bifurcation point where the system jumps back to G2 state is 56.15 kJ/mol (Fig. 3A). $\Delta G$ levels being close to this 'tipping point' or far away from it exhibits different patterns of perturbations (Figs. 3B, C). Under the same stochastic forcing of $z$, system with $\Delta G=56.17$ kJ/mol has a larger variance than with $\Delta G=57.00$ kJ/mol (Fig. 3D), suggesting the increasing risk of turning back into G2 state. Besides, M state with $\Delta G$ =56.17 kJ/mol exhibit stronger autocorrelation (Fig. 3E), which indicates the slower recovery to steady state (28). Figs. 3F and G show that the variance and autocorrelation increases continuously when $\Delta G$ approaching 56.15 kJ/mol. The other branch of bifurcation diagram shows the similar tendency (Fig. S3). These results suggest that the intrinsic noise of *z* reflects the cellular free energy level.

**Antagonistic chemical fluxes increase the robustness and irreversibility of G2/M transition.** We have discussed how the strength of the $\Delta G$ underlies the dynamics of G2/M system. However, the effects of increasing $\Delta G$, or increasing ATP while keeping [ADP] and [Pi] constant, in each individual PdP module, characterized by the directions of the chemical fluxes, are not identical. In the process of Wee1 or Cdc25 phosphorylation, ATP acts as pushing force towards the full phosphorylation of these two regulators. Increasing ATP promotes the inactivation of Wee1 and activation of Cdc25 and consequently facilitates the activation of the central kinase Cdc13/Cdc2. In the central cycle, increasing ATP raises the reaction rate of phosphorylation by Wee1 and thus hinders the Cdc2 activation. We describe the effects of increasing ATP in each cycle as '+'(towards activation of central kinase) or '−'(away from its activation). The 'wildtype' mechanism we discussed above is thus represented by '+ + −'.

To understand the biological significance of the antagonistic fluxes, the first natural question is whether the '+ + −' is the only design of fluxes to give rise to bistable switch behavior. Under the same topology of this network, there exist 8 ($2^3$) designs of fluxes (Fig. S4). Each of these eight are distinguished by the directions of fluxes pumped by ATP hydrolysis. To be general, we analyze their behaviors when increasing *T* or *ΔG* and find that only two of them ('+ + −' and '+ + +') exhibit bistability (Fig. S5). '+ + +' mechanism also has a positive feedback and a double-negative feedback. In fact, all of the rest six have one or two of these two feedbacks replaced by negative feedback, leading to failure of producing bistability. In '+ + +' network, the phosphorylated form of central kinase is its active form (Fig.4A). Contrary to '+ + −', the positive feedback of '+ + +' is coupled with the phosphorylation of central kinase and the double-negative feedback of '+ + +' is coupled with the dephosphorylation process. Consequently, for '+ + +', increasing ATP has the same effects in all three PdP module: promoting the activation of central kinase.

It is intuitively obvious that '+ + −' is less energy-economical than '+ + +' for one flux resisting the other two. What are the benefits for cells to adopt '+ + −' instead of '+ + +' in G2/M circuit? To understand this, we draw the phase diagram of '+ + +'. Both mechanisms have bistable regions, and increasing *T* or $\Delta G$ drives the systems out of such regions (Figs. 4B and C). The slope of upper boundaries, denoted as *s*, characterizes the sensitivity of *T* thresholds to $\Delta G$ changes. With the intrinsic noises of $\Delta G$ in cells considered, high values of *s* might result in the large fluctuations of *T* thresholds. Since cellular G2/M transition is driven by Cdc13 accumulation (increasing *T*), this hypersensitivity might cause the disturbance of cycle period. As is shown in Fig. 4D, in physiological range, increasing $\Delta G$ reduces the sensitivity, and '+ + −' mechanism shows better stability of *s*. For example, ATP decreasing by 5% from $\Delta G$ =56 kJ/mol leads to *T* threshold increasing by 16.4% under '+ + +' but only 7.5% under '+ + −'.

When *T* is kept constant, the width of the bistable region determines the tolerance of M-state Cdc2 activity to $\Delta G$ fluctuations. Rising *T* increases the width of '+ + −' but narrows the bistable range of '+ + +' (Fig. 4E). With T>3, '+ + −' can tolerate $\Delta G$ decreasing over 2 kJ/mol, equivalent to ATP falling by 55.4%, while '+ + +' cannot tolerate ATP falling by as small as 7.8%. This difference gives '+ + −' an additional advantage: better irreversibility of G2/M transition, as is shown in Figs. 4F-H. When the M-state systems is exposed to Gaussian noise of $\Delta G$, '+ + −' is rather robust while '+ + +' mechanism easily switches back to G2 state. This 'falling back' might be very dangerous for cells since it disrupts the order of cell cycle events.

## Discussion

Multisite phosphorylation as a type of post-modification has been found in a great variety of molecular events (26). Intriguing examples include MAPK cascade (29, 30), transcriptional regulation (31), and circadian clock (32) and so on. In this article, based on a simple model describing the multiple PdP cycles catalyzed by enzymes away from zero-order zone, it is shown that free energy level, quantized by phosphorylation potential, governs the response curves to stimuli. Especially, increasing $\Delta G$ strengthens the sensitivity. Compared to a Menten-Machealian steady-state response, an ultrasensitive response has better robustness against noises and helps to amplify the input signal (14, 15, 21, 26). The index $4R_{0.5}/(1-R_{0.5})$ in our model roughly represents the amplification factor of the module. At physiological range of $\Delta G$ with *N*=20, $R_{0.5}$≈0.546 means stimuli increasing by 83% causes the output increasing by four folds. Thus the signal is amplified by 4/0.83≈4.82 folds. Combined with the former theoretical analysis on single PdP cycle operating in zero-order condition (21), our results might suggest a universal trade in cells for amplification in the price of free energy consumption.

G2/M transition in vivo is thought to be propelled by the continuous synthesis of cyclin B (1, 4, 9). It is indicated from our computation that reduced $\Delta G$ elevates the cyclin B threshold and delay the onset on mitosis. This might serves as a 'energy checkpoint' which slow down or even stop those 'energy-insufficient' cells. G1/S transition, another decision making process during cell cycle, is also dominated by multisite phosphorylation (33, 34). The target kinase Clb-CDK is inhibited by Sic1. This inhibition is removed after Sic1 is phosphorylated on at least six sites by a Cln-CDK, which leads to ubiquitination of Sic1 by $SCF^{Cdc4}$, and then its degradation by proteasome (33, 34). The multiple sites give rise to switch-like response of Sic1 to Cln-CDK, which is thought to ensure the proper timing of DNA replication (33, 34).

However, as is shown in our model, $\Delta G$ determines the peak ($y_N^{max}$), the threshold ($x_{half}$), as well as the steepness ($R_{0.5}$) of the response curve. It could be similarly predicted that ΔG fluctuations affects the threshold for Cln-CDK and the timing of entering S phase. Both G1/S and G2/M checkpoint might represent an elaborate design through which cell cycle paces are adjusted to fit the free energy level.

The bifurcation diagram shows that free energy input turns G2/M transition from monostable to bistable, thus ensuring the irreversibility. A comparison between '＋＋−' and '＋ ＋ ＋' suggests a second level of the story. By providing mutually antagonistic chemical fluxes, cellular G2/M circuit becomes more robust to ΔG fluctuations. This adds a new sight towards the evolutionary significance of the G2/M network design.

34. Nash P, et al. (2001). Multisite phosphorylation of a cdk inhibitor sets a threshold for the onset of dna replication. *Nature* 414(6863): 514-521.

# Figures

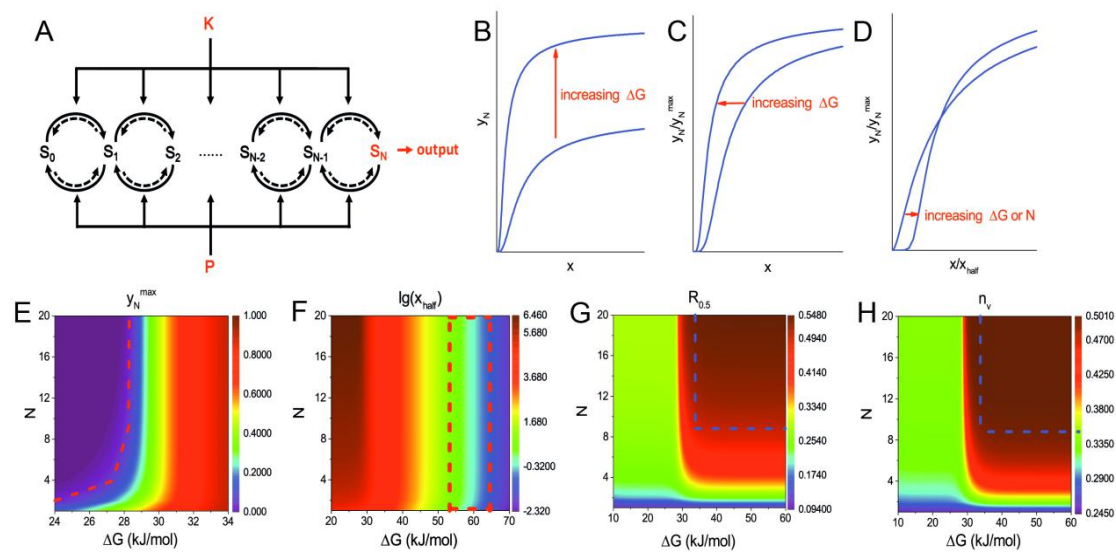

**Fig. 1. Both $N$ and $\Delta G$ governs the responses of PdP cycles to stimuli in a sequential mechanism.** **(A)** Schematic representation of substrate ($S$) multisite phosphorylation catalyzed by kinase ($K$) and phosphatase ($P$). The substrate has $N$ modification sites, thus exists in $N+1$ states. Each state is distinguished by the number of phosphorylated sites, which is marked as the subscript. The solid arrows on PdP cycles represent the directions of enzymatic reactions, as well as the directions of ATP hydrolysis and chemical fluxes. The dashed arrows represent the reverse reactions of catalysis, which is extremely slow compared to the forward reactions. **(B-D)** The steady-state response curves of $y_N$ to $x$ under two different levels of $\Delta G$. In **(C)** and **(D)**, the y-axis is normalized with $y_N^{max}$. Additionally, in **(D)**, the x-axis is normalized with $x_{half}$. **(E-H)** Four indexes regarding the response curves, $y_N^{max}$, $x_{half}$, $R_{0.5}$ and $n_v$ are determined by both $N$ and $\Delta G$. In **(E)**, the red boundary roughly represent the 'dead district' where the module completely loses its function. In **(F)**, the red box represents the range [53.2 kJ/mol, 64.5 kJ/mol] of $\Delta G$, where $x_{half}$ is between 1 and 10. In **(G)** and **(H)**, the blue boxes represents the region where $R_{0.5}$ or $n_v$ becomes rather robust to $N$ and $\Delta G$. Computations are done with $\tilde{b} = \tilde{d}$ and $\tilde{c} = 10^5 \tilde{b}$.

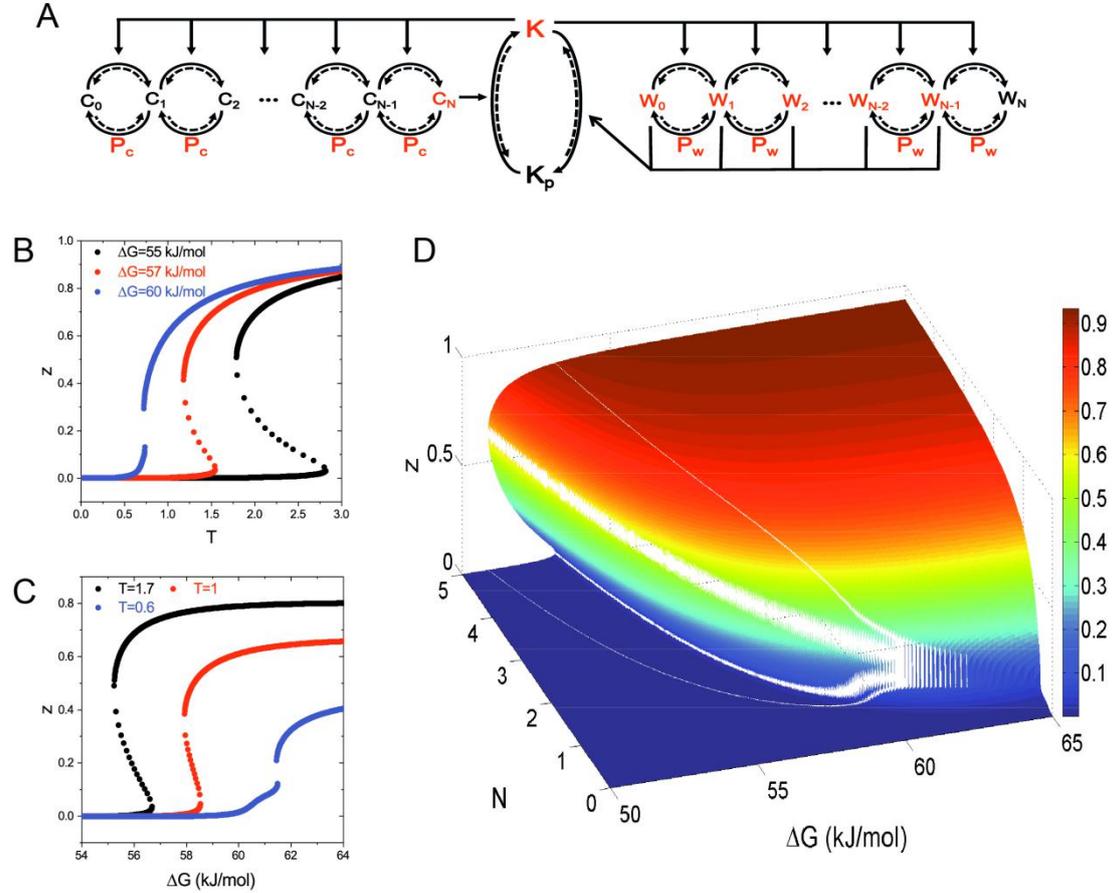

**Fig. 2. $\Delta G$ underlies the bistability of G2/M transition.** (**A**) A simplified representation of the reaction network among $C$ (Cdc25), $K$ (Cdc13/Cdc2) and $W$ (Wee1). The active components are shown in red. (**B**) Cdc13-driving activation of Cdc2 under three different levels of $\Delta G$. Unstable fixed points are shown in dots. With $\Delta G$=55 or 57 kJ/mol, the transition is accompanied by hysteresis, while for $\Delta G$=60 kJ/mol, this transition becomes monostable. (**C**) $\Delta G$-driving activation of Cdc2 with constant levels of $T$. Among these three, $T$=0.6 turns this transition into monostable. (**D**) Three-dimensional phase diagram of G2/M transition shaped by both $T$ and $\Delta G$. The middle layer of the fold surfaces represents the unstable fixed points of the system. Computations are done with $\tilde{k}_{-1} = \tilde{k}_{-2} = 10^{-5}$, $k_{+2} = 1$, $N = 5$, and $\theta = 1$.

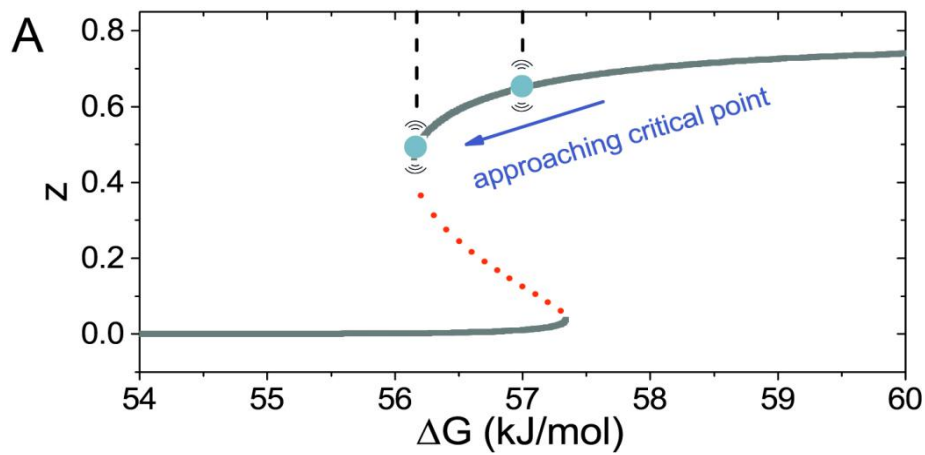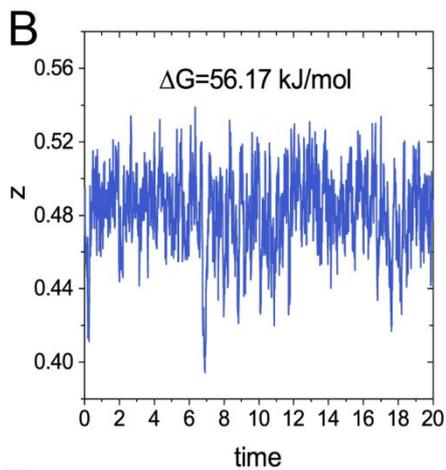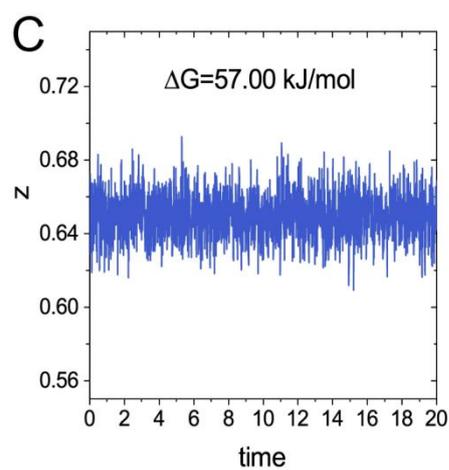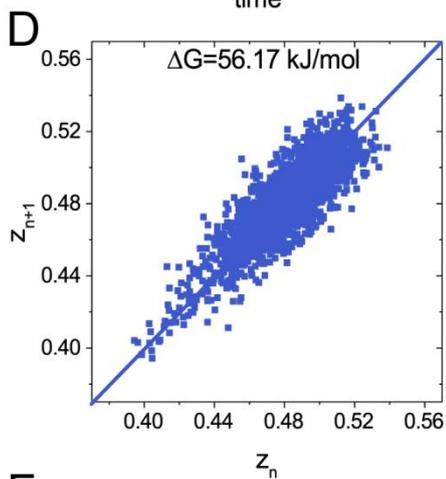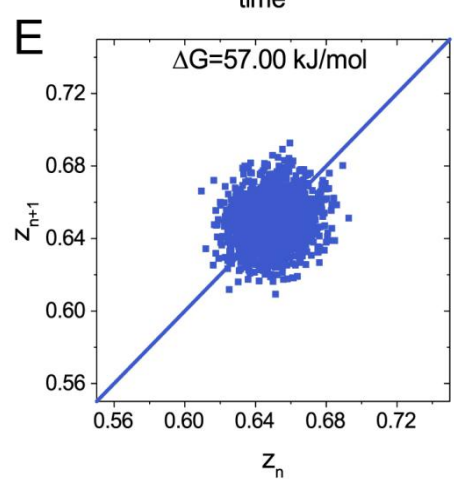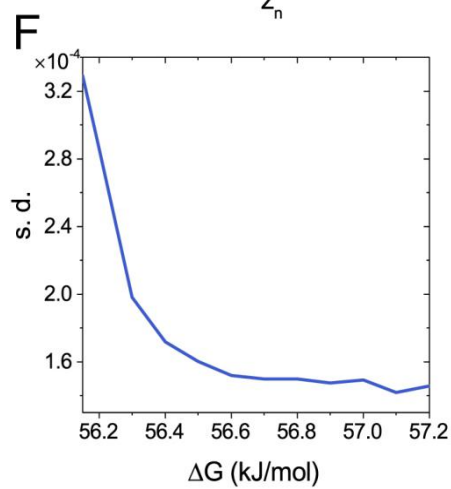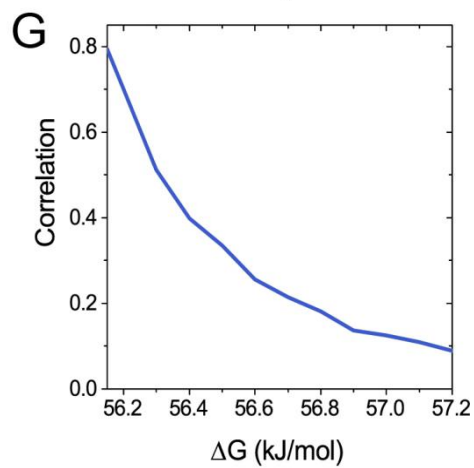

**Fig. 3. Critical slowing down when $\Delta G$ approaches a bifurcation point.** (**A**) Bifurcation diagram of $z$ to $\Delta G$, with $T$=1.4. (**B**) and (**C**) Fluctuation patterns of z when exposed to Gaussian white noises under different levels of $\Delta G$. (**D**) and (**E**) Plots of $z_n$ (the value of $z$ at the *n*th time interval) with $z_{n+1}$ (the value of $z$ at the *n+1*th time interval). (**F**) and (**G**) Standard variation of $z_n$ and correlation coefficient between $z_n$ and $z_{n+1}$ increases continuously when $\Delta G$ approach the critical point. Computations are done with *T*=1.4.

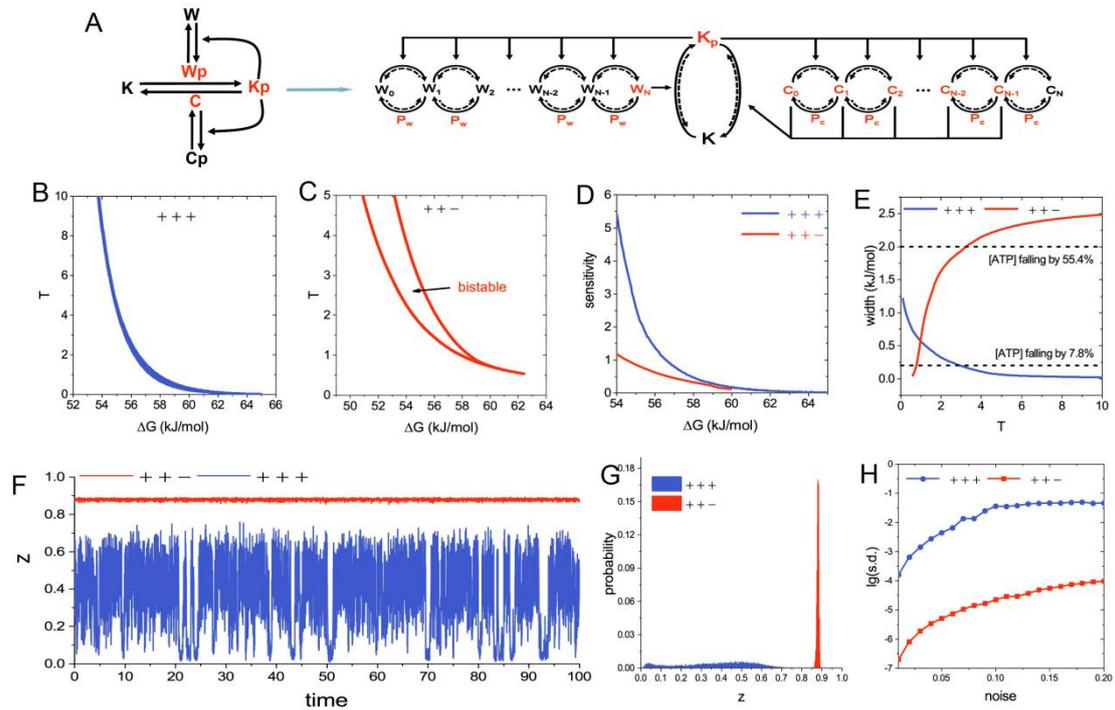

**Fig. 4. Comparisons between '+ + −' and '+ + +' mechanisms.** (**A**) The reaction network of '+ + +' mechanism. The phosphorylated central kinase ($K_p$) is its active form. $K_p$ activate the its activater, Wee1 kinase, while inhibits its inhibitor, Cdc25 phosphatase. (**B**) and (**C**) Phase diagrams of these two mechanisms. (**D**) The sensitivity of T threshold to $\Delta G$, which is quantized by the slope of the upper boundary of bistable region, as a function of $\Delta G$. (**E**) The width of bistable range of $\Delta G$ as a function of $T$. (**F**) and (**G**) After reaching M state, the fluctuations of $z$ when both system are exposed to Gaussian noise of $\gamma = \gamma_0(1 + \lambda \cdot N(0,1))$. $N(0,1)$ represents the standard normal distribution. Computations are done with $T=2$, $N=5$, $\lambda=0.1$. $\gamma_0=2.76\times10^9$ and $4.78\times10^9$ for '+ + −' and '+ + +', respectively. (**H**) The standard deviation of $z$ under different strength of noise ($\lambda$).